\documentclass[]{spie}
\usepackage[]{graphicx}

\def\arcsec{\hbox{$^{\prime\prime}$}}

\title{Astroclimate on Mt. Maidanak Observatory
by AZT--22 1.5m telescope observations} 

\author{Alexander S. Gusev and Boris P. Artamonov
\skiplinehalf
Sternberg Astronomical Institute, Universitetskii Pr. 13, 
Moscow 119992, Russia
}

\authorinfo{Send correspondence to A.S.G. \\
A.S.G.: E-mail: gusev@sai.msu.ru}
 
\begin{document} 
\maketitle 

\begin{abstract}
We present results of Mt. Maidanak Observatory astroclimate study. 
Our data based on AZT--22 1.5m telescope observations in 1996--2005.
\end{abstract}


\section{INTRODUCTION}
\label{sec:intro}

Mount Maidanak Observatory at Ulugh Beg Astronomical Institute
(UBAI) of Uzbek Academy of Sciences located in the South Uzbekistan,
on the slopes of the Baisun range, at an altitude of 2600~m.

{\bf Brief history.} The Sternberg Astronomical Institute (SAI) of Moscow
State University began first studies of the astronomical conditions
at the most promising sites in the Central Asia in the end of the 1960s.
One of the principal criteria used to select the telescope site was that
the peak be isolated. In 1975, the SAI performed measurements of the
astronomical climate at Mt. Maidanak with a two-beam instrument.
Visual estimations showed the seeing 
$\epsilon = 0.6\arcsec$. Later SAI
installed 3 telescopes (1.5m AZT--22 and two smaller instruments) on
the observatory. Currently, all telescopes belong to the Mt. Maidanak
Observatory of the UBAI.

\begin{figure}
\begin{center}
\begin{tabular}{c}
\includegraphics[height=10cm]{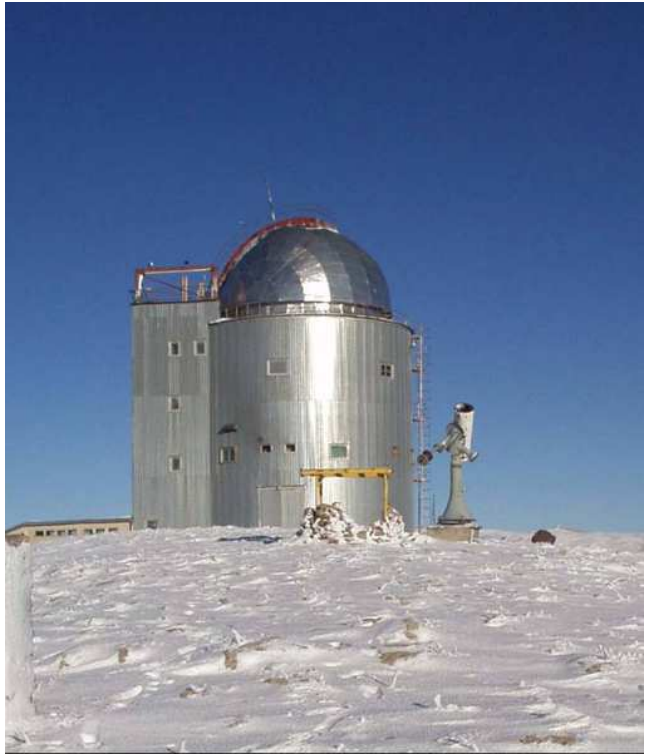}
\end{tabular}
\end{center}
\caption[f1]
{\label{fig:f1}
AZT--22 tower.}
\end{figure}

{\bf 1.5m AZT--22 telescope} (Fig.~\ref{fig:f1}) was designed 
by Leningrad Optics and 
Mechanics Amalgamation (LOMO). It was installed at Mt. Maidanak at the 
end of the 1980s, and made its first light detection in 1991. A two mirror
quasi-Ritchey--Chretien system with a relative aperture of 1:8 is the
principal optical system of the telescope.

{\bf Astroclimate.} The results of extensive studies of the observing
conditions at Mt. Maidanak performed in the 1960s and 1970s were
confirmed in the last decade   with a series of observations of the
astronomical climate using modern techniques. The four-year
observations (1996-99) using the Differential Image Motion Monitor
(DIMM) revealed a mean seeing of $0.69\arcsec$ 
(Ehgamberdiev et al.\cite{Ehgamberdiev00}).
Later studies of the atmosphere at Mt. Maidanak with the MASS
instrument (Kornilov et al.\cite{Kornilov09}) detected a 
free--atmosphere seeing
of $0.47\arcsec$ at heights of 0.5 km and more above the 
level of Mt. Maidanak. Here we are presenting the study of the 
realistic seeing of CCD images obtained on the AZT--22 telescope.

\section{DATA}
\label{sec:data}

To estimate the seeing based on a large sample of CCD observations
obtained at the 1.5m telescope, we used CCD images (21610 frames) of
various astronomical objects in the $U$ (1871), $B$ (3818), $V$ (4369),
$R$ (8415), and $I$ (3137) filters available at the SAI. These
observations were performed in 1996--2005 with CCD cameras.

We estimated the seeing using the sizes of stars in the CCD frames,
taking a star's visible size to be its FWHM profile diameter. For
most of the images we used, the scale is 
$0.2667\arcsec$ per pixel.
Typical exposure times of our observations were 5--300 s.

\begin{figure}
\begin{center}
\begin{tabular}{c}
\includegraphics[height=8cm]{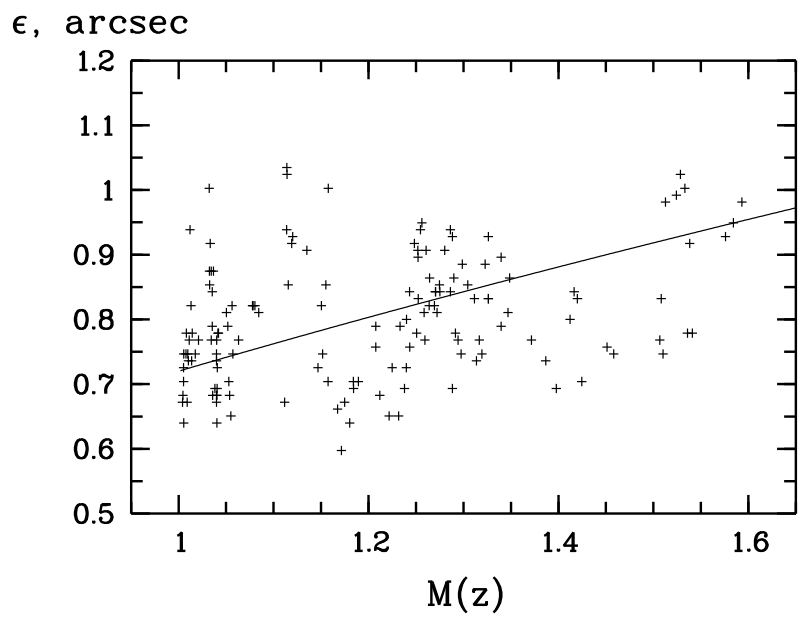}
\end{tabular}
\end{center}
\caption[f2]
{\label{fig:f2}
Seeing $\epsilon$ from $V$ and $R$ CCD images as a function
of air mass $M(z)$ for eight nights with good
$\epsilon_{med}^V \le 0.85\arcsec$, using only 
frames taken no
earlier than 3 hours after sunset. The solid curve is the
$\epsilon(z) \sim M(z)^{3/5}$ relation.}
\end{figure}

\section{RESULTS}
\label{sec:results}

\subsection{CCD seeing}
\label{sec:ccd}

{\bf Dependence of the CCD seeing on exposure time.} Dependence between
the image size and the exposure time due to the telescope drive in
hour angle. To remove the $\epsilon(t)$ dependence, we measured the
sizes of stellar images along the $\delta$, which are independent of
the exposure time.

{\bf Dependence of the CCD seeing on air mass.} 
For seeing $\epsilon < 0.9\arcsec$ and air masses
$M(z) \equiv \sec z < 1.6$, there is a known relation for the seeing:
$\epsilon(z)$ (Sarazin and Roddier\cite{Sarazin90}):
\begin{equation}
\label{eq:eq1}
\epsilon(z) =  \epsilon(0) M(z)^{3/5}.
\end{equation}
For the data shown in Fig.~\ref{fig:f2}, 
$\epsilon(0) = 0.72\arcsec$.

\begin{figure}
\begin{center}
\begin{tabular}{c}
\includegraphics[height=15cm]{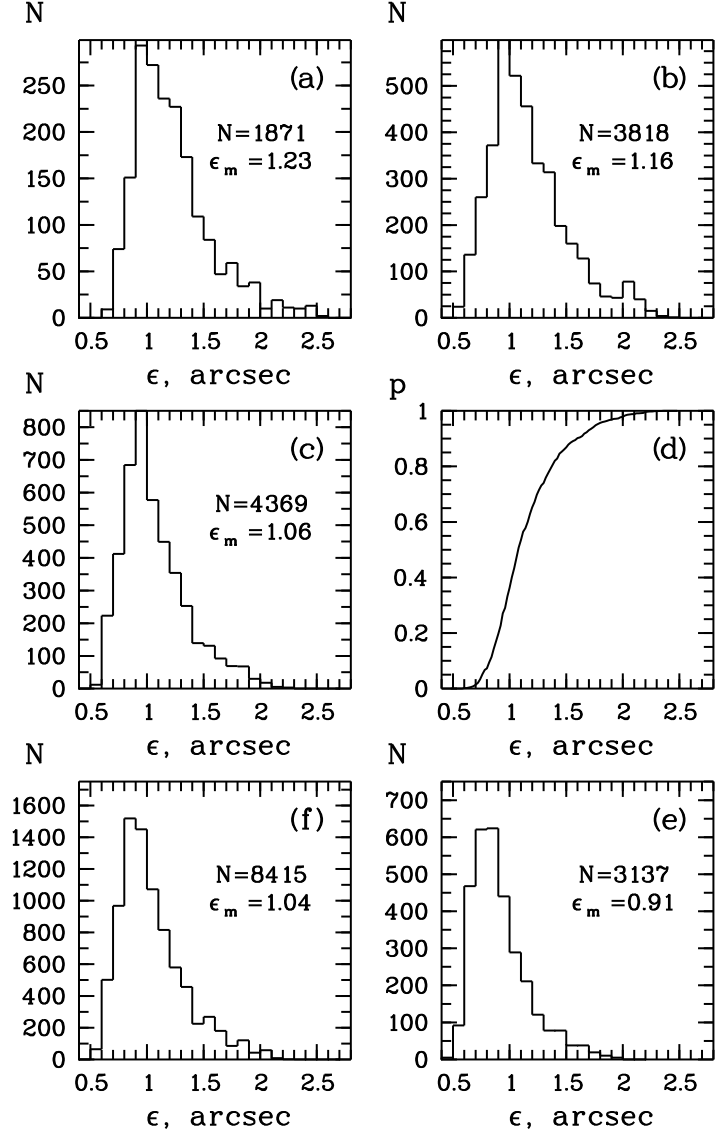}
\end{tabular}
\end{center}
\caption[f3]
{\label{fig:f3}
Distributions of the seeing $\epsilon$ measured in the
{\bf (a)} $U$, {\bf (b)} $B$, {\bf (c)} $V$, {\bf (e)} $R$, and
{\bf (f)} $I$ bands and {\bf (d)} the integrated seeing distribution
in the $V$ band. The total number of frames $N$ and the median
seeing $\epsilon_{med}$ are indicated for each band.}
\end{figure}

{\bf Seeing from CCD images in various filters.} Fig.~\ref{fig:f3} 
presents histograms of the seeing measured in the $U$, $B$, $V$, $R$,
and $I$ bands, as well as the integrated seeing in the $V$ band. The
median seeing, $\epsilon_{med}$, ranges from 
$1.23\arcsec$ in $U$ to
$0.91\arcsec$ in $I$. The seeing in $V$ varies from 
$0.5\arcsec$ to $2.4\arcsec$, with 
$\epsilon_{med}^V = 1.065\arcsec$. The number of
images with $\epsilon^V \le 0.7\arcsec$ does not 
exceed $2\%$, and the number with 
$\epsilon^V \le 0.8\arcsec$ does not exceed $8\%$ 
(Fig.~\ref{fig:f3}d).

The derived median seeing values in the various photometric bands
are in good agreement with the following relation between $\epsilon$
and $\lambda$:
\begin{equation}
\label{eq:eq2}
\epsilon(\lambda) = (10.52 \pm 0.03)\lambda^{-1/5} + const,
\end{equation}
where $\lambda$ is the wavelength (in \AA) for the maximum transparency of
the corresponding filter.

Using the median values $M(z)_{med} = 1.22$ and
$\epsilon_{med}^V = 1.065\arcsec$ for the AZT--22 frames,
we estimated the characteristic seeing reduced to unit air mass,
$\epsilon_{med}^V(M(z)=1)$. According to Eq.~(\ref{eq:eq1}),
$\epsilon_{med}^V(M(z)=1) = 0.945\arcsec$. Since, 
strictly speaking,
Eq.~(\ref{eq:eq1}) is valid only for the free atmosphere, our estimate of
$\epsilon_{med}^V(M(z)=1)$ should be treated cautiously.

\begin{figure}
\begin{center}
\begin{tabular}{c}
\includegraphics[height=5cm]{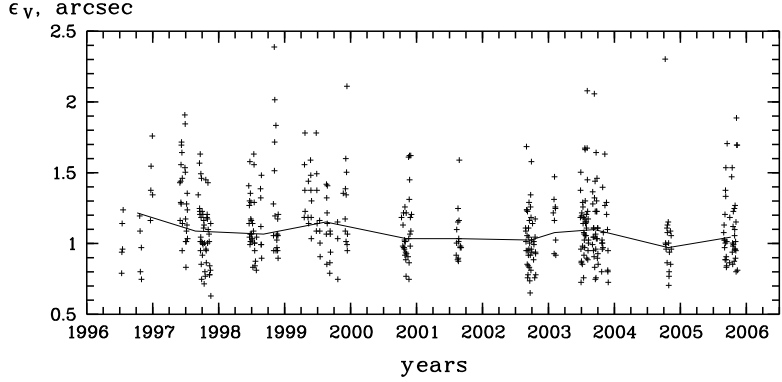}
\end{tabular}
\end{center}
\caption[f4]
{\label{fig:f4}
Nightly (crosses) and yearly (broken line) averages
of the $V$-band seeing $\epsilon$.}
\end{figure}

\begin{table}[h]
\caption{Distribution of CCD seeing $\epsilon$ over months.}
\label{tab:t1}
\begin{center}
\begin{tabular}{|l|c|c|c|c|}
\hline
\rule[-1ex]{0pt}{3.5ex}   & Number of & Number of & Total & \\
\rule[-1ex]{0pt}{3.5ex} Month & observation & observation & 
number of & $\epsilon_{med}^V\arcsec$ \\
\rule[-1ex]{0pt}{3.5ex}   & years & nights & frames &  \\
\hline
\rule[-1ex]{0pt}{3.5ex} January   & 1 &   2 &    59 & 1.29 \\
\rule[-1ex]{0pt}{3.5ex} February  & 1 &   7 &   459 & 1.08 \\
\rule[-1ex]{0pt}{3.5ex} March     & 0 &   0 &     0 &  --  \\
\rule[-1ex]{0pt}{3.5ex} April     & 1 &   5 &   213 & 1.44 \\
\rule[-1ex]{0pt}{3.5ex} May       & 2 &  10 &   399 & 1.32 \\
\rule[-1ex]{0pt}{3.5ex} June      & 4 &  38 &  1424 & 1.34 \\
\rule[-1ex]{0pt}{3.5ex} July      & 5 &  63 &  1838 & 1.08 \\
\rule[-1ex]{0pt}{3.5ex} August    & 6 &  44 &  1366 & 1.10 \\
\rule[-1ex]{0pt}{3.5ex} September & 6 &  93 &  5542 & 1.06 \\
\rule[-1ex]{0pt}{3.5ex} October   & 9 &  93 &  5186 & 1.01 \\
\rule[-1ex]{0pt}{3.5ex} November  & 7 &  68 &  4024 & 1.04 \\
\rule[-1ex]{0pt}{3.5ex} December  & 3 &  17 &  1100 & 1.29 \\
\hline
\rule[-1ex]{0pt}{3.5ex} Total    & 10 & 440 & 21610 & 1.06 \\
\hline
\end{tabular}
\end{center}
\end{table}

\begin{figure}
\begin{center}
\begin{tabular}{c}
\includegraphics[height=11cm]{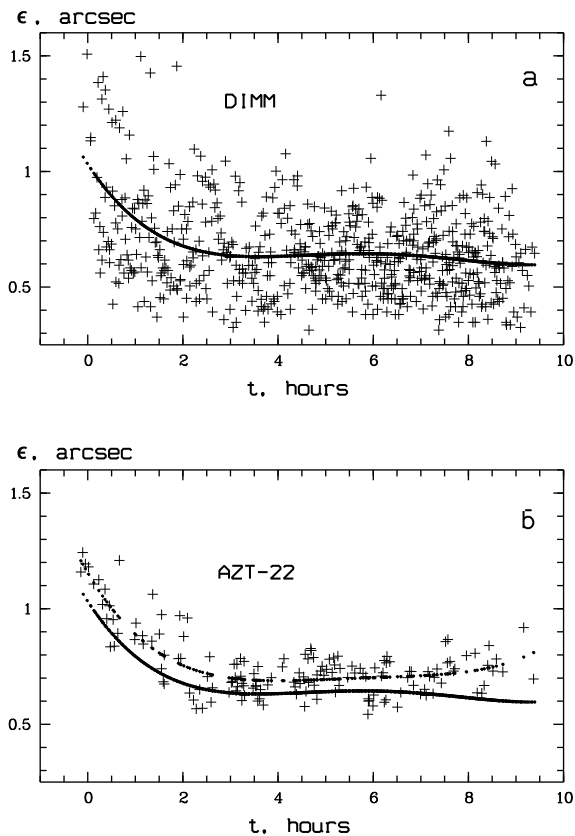}
\end{tabular}
\end{center}
\caption[f5]
{\label{fig:f5}
Comparison of the seeing during the course of a night
measured {\bf (a)} with the DIMM and {\bf (b)} at the AZT--22. DIMM
data for eight nights with good $\epsilon$ values averaged with a
fourth-power polynomial are shown together with the
$V$ and $R$ AZT-22 data ($\epsilon_V$ and $\epsilon_R$, reduced to unit
air mass) for the same nights, averaged with a 4th-power polynomial.
The time 0 hours correspond to the end of astronomical twilight.}
\end{figure}

{\bf CCD seeing over years.} Fig.~\ref{fig:f4} displays the 
$V$-band seeing for each
observing night (nightly median $\epsilon_V$ values are plotted). In
total, we reduced data for 440 nights between July 1996 and November
2005.

The scatter of the nightly mean $\epsilon_V$ values is fairly large:
from $0.6\arcsec$ to $2.3\arcsec$, 
but the seeing is $0.8\arcsec$ -- 
$1.4\arcsec$ on most nights. During the entire
observation period, $\epsilon_V$ was better than
$0.7\arcsec$ on only two nights ($0.5\%$) and better than 
$0.8\arcsec$ on 31 nights ($7\%$).

To estimate long-term seeing trends, we calculated the median
$\epsilon_V$ values for each observation year.

Excluding the data for 1996 (a year with poor statistics, with
only 15 observation nights), we find that the seeing did not change
during the studied period, on average. The range of $\epsilon_V$ was
$0.97\arcsec$ (2005) -- $1.15\arcsec$ (1999).

{\bf CCD seeing over months.} The best seeing is achieved in Autumn 
(with the median $\epsilon_V = 1.01\arcsec$ for October 
observations), while the worst seeing is obtained in the 
Winter--Spring (with $\epsilon_V$ reaching $1.44\arcsec$ 
in April). Table~\ref{tab:t1} shows the months 
seeing statistics. Our data generally reproduces the results of
Ehgamberdiev et al.\cite{Ehgamberdiev00}.

\subsection{Time for relaxation to stationary conditions in the 
free atmosphere and under the AZT--22 dome}
\label{sec:relax}

To estimate the time for relaxation to stationary atmospheric
conditions under the AZT--22 dome with the ventilation on, we
analyzed the variations of the CCD $V$ and $R$ seeing during the
course of a night for eight nights with
$\epsilon_{med}^V \le 0.85\arcsec$. Note, here we 
used the $\epsilon$
values reduced to unit air mass in accordance with Eq.~(\ref{eq:eq1}). 
The results show that the time scale for relaxation to 
stationary conditions under
the dome is 2--2.5 hours after the end of astronomical twilight
(Fig.~\ref{fig:f5}b). Fig.~\ref{fig:f5}a shows the seeing measured on 
the same nights with the DIMM (Ehgamberdiev et al.\cite{Ehgamberdiev00}) 
mounted near the AZT--22 tower.

Simultaneous seeing estimates with the DIMM and at the AZT--22
can be used to estimate the influence of the space under the
telescope dome. The seeing in the free atmosphere differs from that
under the dome by about $0.1\arcsec$ when the 
ventilation is on (Fig.~\ref{fig:f5}a). A powerful ventilation system is 
able to create a laminar flow and to improve the seeing to within 15 minutes 
for similar telescopes and dome construction 
(Artamonov et al.,\cite{Artamonov10}).
The AZT--22 ventilation system is probably not strong enough to
provide an appreciable improvement in the seeing. The relaxation
of the thermal conditions under the AZT--22 dome takes approximately
the same time as in the free atmosphere. The difference between the
DIMM and AZT--22 estimates remains constant, and this remains
unexplained for the moment.

\acknowledgments

This study was supported by the Russian Foundation
for Basic Research (project nos. 08--02--01323, 09--02--00244, 
and 10--02--91338).

\bibliography{gusev}
\bibliographystyle{spiebib}   

\end{document}